# A Novel Approach to Chest X-ray Lung Segmentation Using U-net and Modified Convolutional Block Attention Module

Mohammad Ali Labbaf Khaniki*, Marzieh Mirzaeibonehkhater**, Mohammad Manthouri***, Katayoon Faraji****

* Faculty of Electrical Engineering, K.N. Toosi University of Technology, Tehran, Iran

** Department of Electrical and Computer Engineering, Indiana University-Purdue University

*** Faculty of Electrical and Electronic Engineering, Shahed university, Tehran, Iran

**** Martin J. Whitman School of Management, Syracuse University, Syracuse, NY, USA

mohamad95labafkh@gmail.com*

marzieh89mirzaei@gmail.com**

mmanthouri@shahed.ac.ir***

kfaraji@syr.edu****

**Abstract:**

Lung segmentation in chest X-ray (CXR) images is a crucial task in medical image analysis, aiding in accurate disease diagnosis and treatment planning. This study presents an improved U-Net architecture by integrating a Convolutional Block Attention Module (CBAM) to enhance segmentation performance. The proposed CBAM unifies three attention mechanisms (Channel Attention, Spatial Attention, and Pixel Attention) to refine feature extraction and improve focus on critical image regions. The Channel Attention mechanism emphasizes inter-channel dependencies, the Spatial Attention mechanism enhances localization accuracy by highlighting spatial correlations, and the Pixel Attention mechanism refines segmentation precision at the pixel level. By incorporating CBAM into the U-Net framework, the model achieves superior performance in lung segmentation, as evaluated using the Dice coefficient and Jaccard similarity index. Experimental results demonstrate that the proposed approach significantly improves segmentation accuracy, making it a promising advancement in medical image processing for lung disease assessment.

## 1) Introduction

The analysis of chest X-ray images is a multifaceted and labor-intensive process, frequently requiring the detection of several irregularities at once. This procedure is usually carried out by radiologists manually, which can put a substantial burden on healthcare resources. The intricate nature of the thoracic background in chest X-ray images and the subjective nature of the interpretation can lead to bias and discrepancies in the diagnostic outcomes [1]. Additionally, the quality and resolution of the images, coupled with challenges related to data, can further complicate the interpretation process. Computer-aided detection (CAD) are technological systems that provide support to doctors in interpreting medical images [2]. They process digital images or videos, identify typical patterns, and highlight areas of interest, such as potential diseases, to aid in decision-making. CAD systems are a blend of artificial intelligence and computer vision technologies, combined with radiological and pathology image processing [3]. They are commonly used for tasks like tumor detection. For example, many hospitals employ CAD for routine medical screenings in mammography (for breast cancer detection), colonoscopy for polyp detection, and lung cancer detection. Within CAD systems, segmentation is a key process that precisely distinguishes and separates regions of interest, like tumors or other irregularities, from the rest of the healthy tissue. This significantly improves the accuracy of further analyses, such as determining the size of a tumor or evaluating the progression of a disease [4].

Machine Learning (ML), a branch of Artificial Intelligence (AI), equips computers with the ability to learn from data without the need for explicit programming. It has the capacity to self-adjust and enhance its performance over time with little to no human intervention. ML algorithms are capable of recognizing patterns, multi-objective algorithms [5], and making forecasts [6]. Deep learning has been received many attentions in various fields such as defect detection [7], virtual reality [8] and [9]. Advancements in technology drive the growing necessity to integrate cutting-edge methods AI and ML into image classification and segmentation. [10] explores the interpretability of deep-learned error-correcting codes, analyzing their structure, decoding strategies, and potential advantages over traditional coding methods. This research [11] highlights the effectiveness of deep learning models combined with gray level enhancement techniques for automating the classification of white matter lesions in MS patients' MRI scan. Image processing using deep learning techniques in humanoid robots involves applying neural networks to analyze and manipulate visual data captured by robot cameras, enabling tasks such as object recognition, navigation, and interaction with the environment. [12] evaluates and compares the performance of MedSAM1 and MedSAM2 models for automating left atrial segmentation in 3D late gadolinium-enhanced MRI, demonstrating the potential of these foundational deep learning models in medical image analysis. [13] introduces G-CLAHE (Global-Contrast Limited Adaptive Histogram Equalization), a novel approach that combines global and local contrast enhancement techniques to improve the quality and diagnostic accuracy of medical X-ray images. This study [14] reveals that large-scale folds in the bladder dome, rather than previously hypothesized small mucosal rugae, are responsible for the organ's high compliance during initial filling and contribute to its remarkable voiding efficiency of over 90%.

Deep learning has brought about a significant transformation in the domain of image segmentation, introducing a multitude of techniques that have substantially enhanced the precision and efficiency of the segmentation process. One such technique is U-Net, a convolutional neural network specifically engineered for the segmentation of biomedical images. It possesses a U-shaped architecture comprising a contracting path (encoder) and an expansive path (decoder). This unique structure enables U-Net to operate with a smaller number of training images while delivering highly accurate segmentation [15]. Unet++ architecture aims to improve medical image segmentation by incorporating nested, dense skip pathways that connect the encoder and decoder sub-networks. These redesigned skip pathways help reduce the semantic gap between the feature maps of the encoder and decoder, which can facilitate the learning process for the optimizer [16]. [17] enhances the original U-Net structure with residual blocks to overcome the vanishing gradient issue and support deeper network models. [18] presented a 3D U-Net architecture for segmenting lung tumors in CT scans and X-Ray images.

Attention mechanisms, which are inspired by the human visual system, have proven to be highly effective in a variety of visual tasks within the realm of image processing and natural language processing [19]. The attention mechanism empowers the model to concentrate on various segments of the input sequence while generating an output. This effectively encapsulates the relationships between words or events, even when they are distantly placed [20]. [21] surveys the role of positional encoding in transformer-based time series models, highlighting various encoding methods, their effectiveness, and open challenges in the field.

They have found applications in numerous tasks such as image classification, object detection, semantic segmentation, video comprehension, image generation, 3D vision, multi-modal tasks, and self-supervised learning. This research [22] enhances face recognition accuracy by intelligently combining features from multiple models. It achieves this by leveraging attention mechanisms and applying information bottleneck principles. The study [23] aimed to evaluate the hazardous aspects of an abandoned gypsum mine by exploring its location, depth, extension, and overall condition using x-ray diffraction technique. [24] introduces an innovative pedestrian detection system that operates in a single stage and incorporates both channel and spatial attention mechanisms within the structure of a CNN. The researchers introduce a method called CACNN, a deep learning approach that skillfully merges 2-D and 3-D CNNs to simultaneously derive spectral and spatial information from hyperspectral images in [25]. The authors propose an end-to-end deep learning model that integrates a multi-branch spectral-temporal CNN with an efficient channel attention mechanism and the LightGBM algorithm in [26]. [27] proposed a U-Net architecture enhanced with multiple encoders for better feature extraction and an attention mechanism in decoders for precise focus on relevant features. [28] improves upon the U-Net model by incorporating multi-scale spatial attention and dilated convolutions to capture context information effectively.

The study introduces a novel method for lung segmentation in chest X-ray images by combining U-net with attention mechanisms. This method improves the U-net structure by integrating a Convolutional Block Attention Module (CBAM), which unifies three separate attention mechanisms: Channel Attention Map, Spatial Attention Map, and Pixel Attention.

Here are the key advancements of this research:

- **Enhanced U-net Architecture**: By adding CBAM to standard convolution layers in the U-net architecture, the model can better capture global contextual information and improve the network's attention to specific regions. This leads to improved feature representation and, ultimately, better performance in tasks such as image segmentation
- **Proposed CBAM with Triple Attention Mechanisms**: The integration of Channel, Spatial, and Pixel Attention mechanisms significantly enhances the model's focus on relevant features within X-ray images. The channel attention mechanism emphasizes inter-channel associations, enabling the model to concentrate more on informative channels and boosting its capability to identify relevant features. The spatial attention mechanism allows the model to focus on crucial spatial locations, enhancing its precision in localization by paying more attention to the spatial correlations between features. Lastly, the pixel attention mechanism empowers the model to focus on individual pixels, further sharpening its focus and improving the accuracy of segmentation by paying more attention to the most informative pixels.

The adoption of the suggested CBAM in conjunction with the U-net architecture marks a significant progression in the field of medical imaging. This innovation is anticipated to improve the precision of diagnoses and contribute to more favorable health outcomes for patients. The method's enhanced performance is assessed using sophisticated metrics, focusing particularly on Dice coefficient and Jaccard similarity indices. These metrics are essential for assessing the segmentation's accuracy by comparing the predicted segmentation with the actual anatomical boundaries, as demonstrated in the studies by [29].

The structure of the manuscript is organized in the following manner: Section II offers a foundational overview of the dataset utilized for Chest X-ray lung segmentation and the methods employed for preprocessing. Section III delves into the methodology being proposed, detailing the fusion of proposed CBAM with the U-net framework. Section IV is dedicated to the exposition of the simulation outcomes, encompassing a comprehensive account of the training and validation procedures of the method under consideration. The paper culminates with Section V, which encapsulates the principal findings and the pivotal contributions put forth by this study.

## 2) Chest X-ray Lung Segmentation Dataset Description

### 2-1) Dataset Description

Researchers used chest X-rays and matching lung masks from Kaggle to train models that could identify lungs in X-rays [30]. The dataset under consideration constitutes a significant asset for the medical research community, especially in the realm of automated image analysis for the screening of tuberculosis. It encompasses a collection of chest X-ray images paired with segmentation masks, albeit with the caveat that some masks may be absent. It is recommended that users corroborate the availability of masks corresponding to the X-ray images to maintain research integrity. The dataset is comprehensive, comprising 360 normal X-ray images and 344 abnormal X-ray images indicative of lung infections. The images have been carefully labeled by experienced radiologists. Representative X-ray images and their corresponding masks from both the training and validation sets are shown in Figure 1.

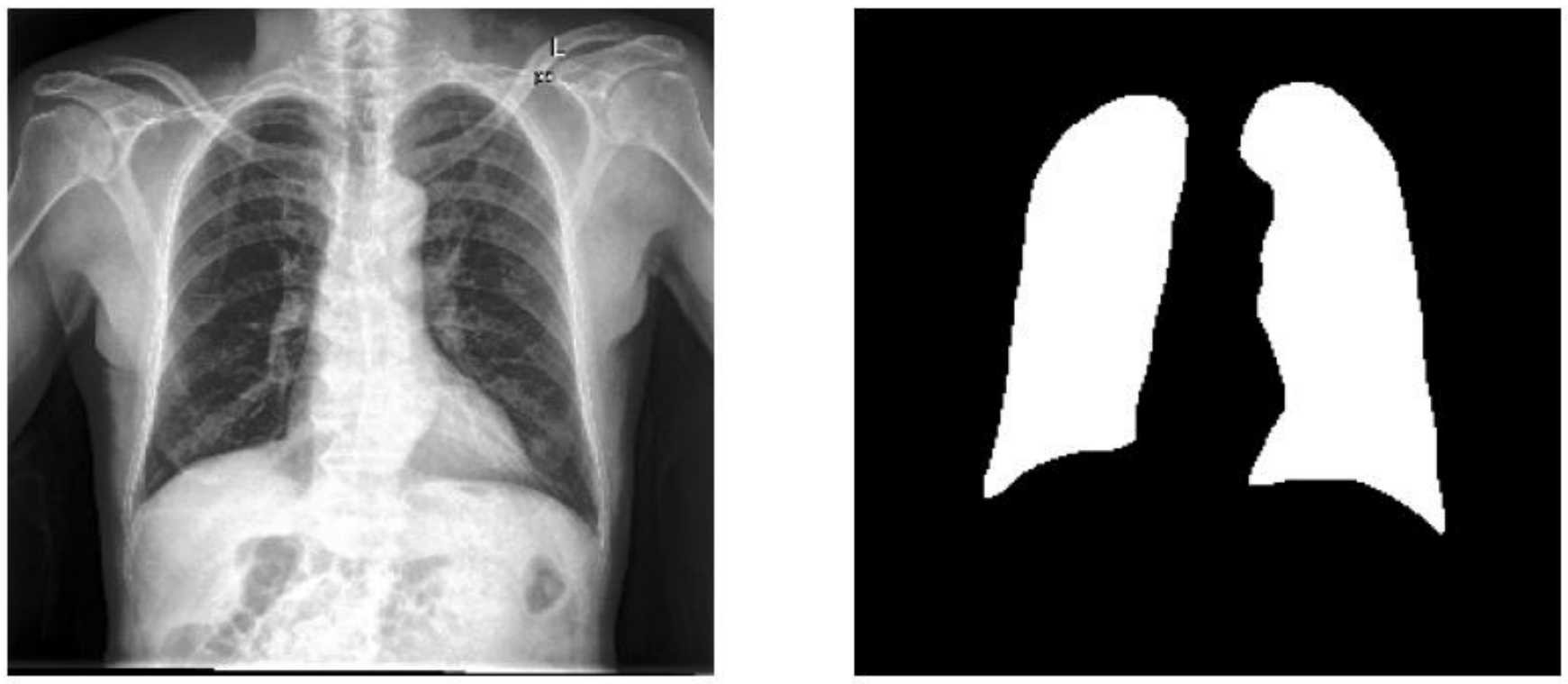

(a)

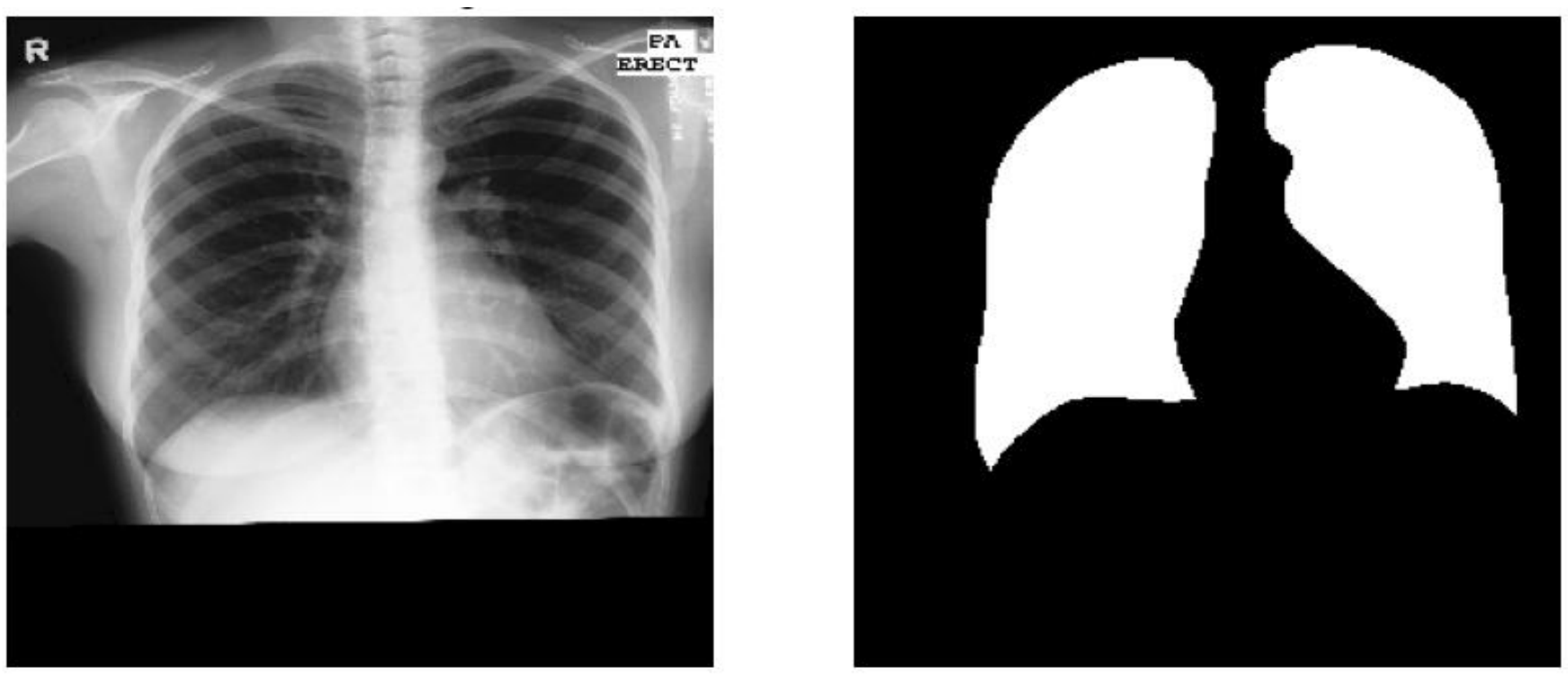

(b)

Fig. 1: Representative X-ray images with their corresponding masks from the training and validation datasets, as annotated by expert radiologists.

The images are rendered anonymous and are presented in the Digital Imaging and Communications in Medicine (DICOM) format, which is the universally accepted standard for the management, storage, reproduction, and dissemination of medical imaging information. The dataset encapsulates a broad array of pulmonary abnormalities, not limited to effusions and miliary patterns. The dataset is a crucial resource for developing algorithms that can detect and segment lung diseases in chest X-rays. It includes a diverse range of images, both normal and abnormal, providing a comprehensive dataset for analysis. This collection is more than just images; it's an essential link between medical knowledge and AI technology, driving progress in automated diagnostics. The meticulous assembly and preparation of the data render it an indispensable asset for researchers striving to innovate in medical image analysis.

**2-2) Image Augmentation and Preprocessing**

In the study, a series of data augmentation techniques were systematically applied to enhance the dataset for training the neural network, resulting in a sixfold increase in the dataset size. In the realm of neural network training for medical image analysis, the augmentation techniques of contrast adjustment and Gaussian blur play pivotal roles. Enhancing the contrast of images serves to amplify the distinction between features, aiding the network in segmenting regions of interest more effectively, and equipping it to handle variations in lighting conditions that are common in clinical environments. Concurrently, the application of Gaussian blur mitigates high-frequency noise, which could otherwise lead to overfitting, and fosters a level of generalization that enables the network to recognize features within images that may lack sharpness or have undergone different preprocessing steps. These techniques collectively ensure that the neural network is not only more robust but also more adept at accurately analyzing medical images under a wide array of conditions. The augmentation process commenced with a contrast adjustment, where the contrast of each image was intensified using a linear formula, thereby generating a set of contrast-enhanced images. Subsequently, a Gaussian blur was applied to these images, introducing a slight blur effect without additional smoothing. This step produced a collection of blurred images.

To further augment the dataset and bolster the robustness of the neural network, a horizontal flipping technique was employed. Horizontal flipping is an essential data augmentation technique that enhances the robustness of neural networks by introducing additional variability and symmetry to the training dataset. By mirroring images along the vertical axis, it ensures that the network learns to recognize and segment features irrespective of their lateral orientation, thus preventing bias towards anomalies that may appear on one side of the body. This method is particularly beneficial in medical imaging, where the accurate detection and segmentation of features are critical, and the mirrored defects aid in reinforcing the network's diagnostic consistency across varying patient positions and imaging angles. The same augmentation procedures were then applied to these horizontally flipped images, resulting in an additional set of contrast-adjusted and blurred images.

The culmination of these augmentation steps not only expanded the dataset but also introduced a variety of transformations that simulate different imaging conditions. This diversity is crucial for training a neural network that is resilient and capable of accurately segmenting lung regions in chest X-ray images under various scenarios. Fig. 2 displays the enhanced images that have been augmented using the specified technique, along with their respective masks.

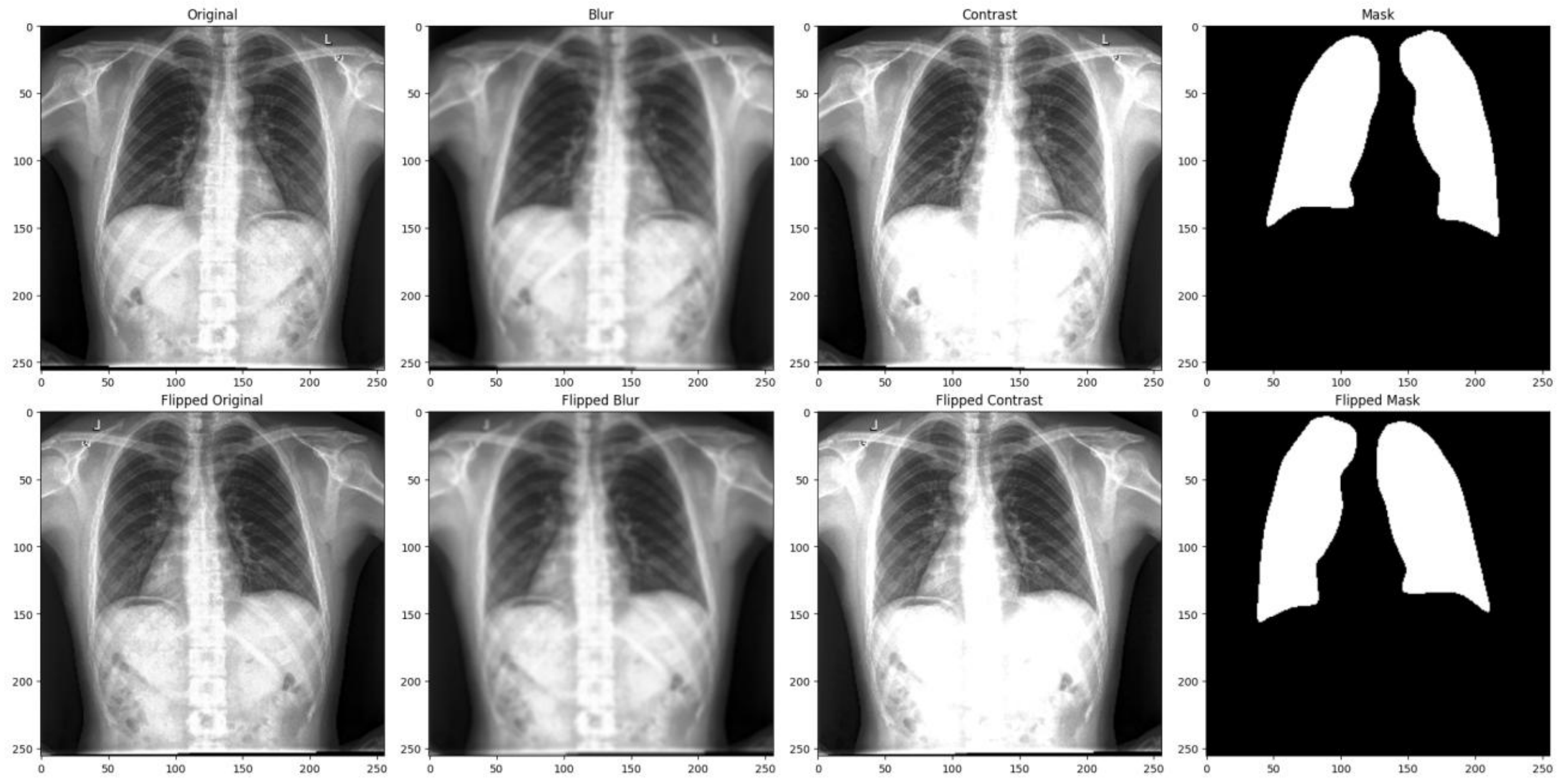

Fig. 2: Enhanced and augmented images with their corresponding masks, utilizing the specified augmentation technique.

## 3) Methodology

In this section, we commence with an introduction to the U-Net architecture tailored for segmenting the lung regions from radiographic images. Subsequently, we present the CBAM, followed by a proposal for an enhanced U-Net framework that integrates the CBAM, aiming to refine the segmentation process.

### 3-1) U-Net architecture

The U-Net architecture is a convolutional neural network designed for biomedical image segmentation. It features a symmetric encoder-decoder structure, which is why it's called "U-Net." The encoder part compresses the image into a feature-rich representation, reducing spatial dimensions while increasing feature complexity through convolution and pooling operations. The decoder part then expands this representation, using transposed convolutions to increase spatial dimensions for precise localization. Skip connections between the encoder and decoder transfer context information, aiding in accurate segmentation. This combination of context from the encoder and localization from the decoder makes U-Net particularly effective for medical imaging tasks. Fig. 3 illustrates the U-Net architecture for lung segmentation.

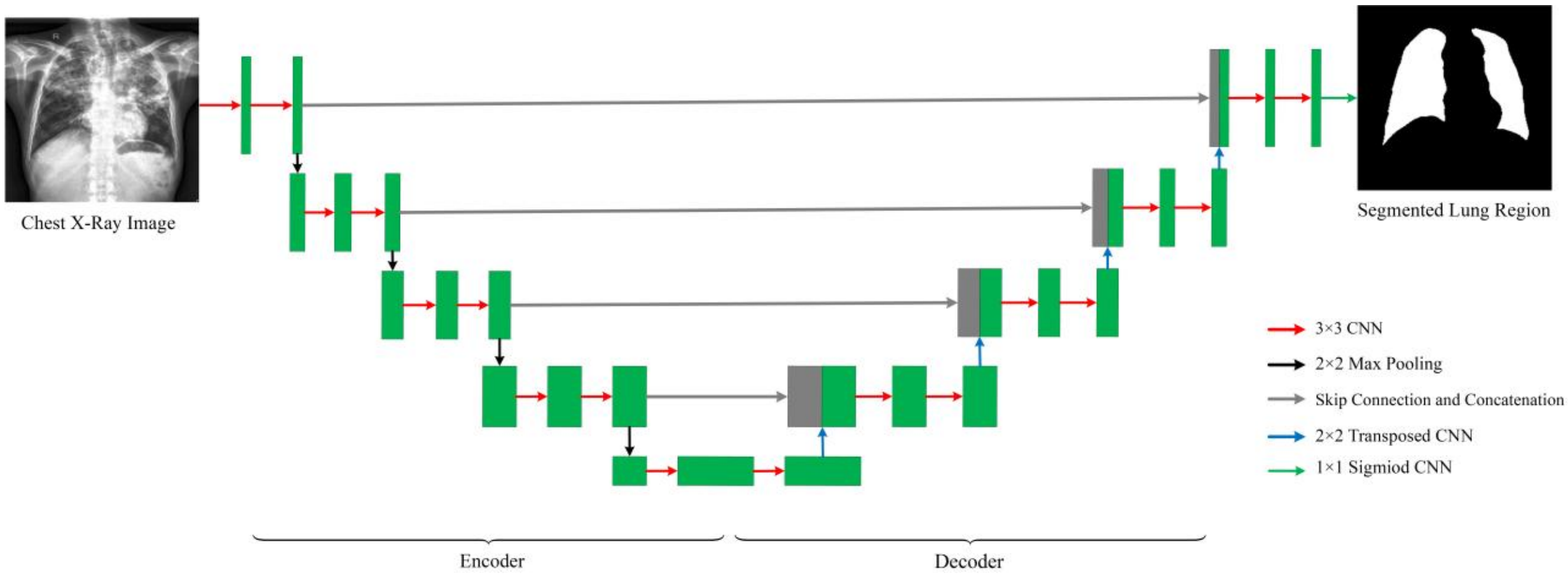

Fig. 3: U-Net architecture applied for lung segmentation using chest X-ray imagery.

### 3-2) CBAM Model

The CBSM is a model designed to enhance the precision of chest X-ray lung segmentation in the U-net architecture, as opposed to the conventional CNN, particularly when dealing with limited sample sizes. The amalgamation of channel, spatial, and pixel attention mechanisms marks a substantial progression in honing the focus on pertinent features within X-ray images.

1. **Channel Attention**: This mechanism emphasizes inter-channel associations, enabling the model to concentrate more on channels that are more informative. It aids the model in focusing on the most significant features across various channels, thereby boosting the model's capability to identify relevant features.

2. **Spatial Attention**: This mechanism allows the model to concentrate on crucial spatial locations. It enables the model to pay more attention to the spatial correlations between features, thereby enhancing the model's precision in localization.

3. **Pixel Attention**: This mechanism empowers the model to focus on individual pixels. It further sharpens the model's focus, enabling it to pay more attention to the most informative pixels, thereby improving the accuracy of segmentation.

These three attention mechanisms collaborate to provide a more comprehensive representation of features, thereby enhancing the model's performance in tasks such as image segmentation. They enable the model to capture global contextual information more effectively and improve the network's attention to specific regions.

Consider a feature map represented by $F \in \mathbb{R}^{H\times W\times C}$, where H and W are the height and width of the input image, respectively, and C represents the count of spectral bands in the input map. The CBSM functions as a method for dynamically adjusting weights to identify important areas within complex scenes. This is achieved through a 1D channel attention map, denoted by $M_C \in \mathbb{R}^{1\times 1\times C}$, a 2D spatial attention map denoted by $M_S \in \mathbb{R}^{H\times W\times 1}$, and a 2D pixel attention map denoted by $M_P \in \mathbb{R}^{H\times W\times 1}$. Throughout the operation of CBSM, the input data undergo a refinement process first through $M_C$, then $M_S$, and finally $M_P$. Consequently, the overall process of the improved CBSM can be written as follows:

The first feature map refined by the channel attention map:

$$F_C = (M_C(F) + 1) \times F. \tag{1}$$

The intermediate feature map further refined by the spatial attention map:

$$F_S = (M_S(F_C) + 1) \times F_C. \tag{2}$$

The final feature map refined by the pixel attention mechanism:

$$F_P = (M_P(F_S) + 1) \times F_S, \quad (3)$$

where $\times$ represents element-wise multiplication, and $+$ denotes element-wise addition, the refined maps $M_C$, $M_S$, and $M_P$ are broadcasted to match the dimensions of the feature maps to which they are applied. The final output, $F_P$, is the feature map that has been sequentially refined by the channel, spatial, and pixel attention mechanisms. This provides a more focused and detailed representation for chest X-ray lung segmentation. This process allows for more granular control over the attention given to each pixel, potentially improving the accuracy of the segmentation. The CBSM is visualized in Figure 4.

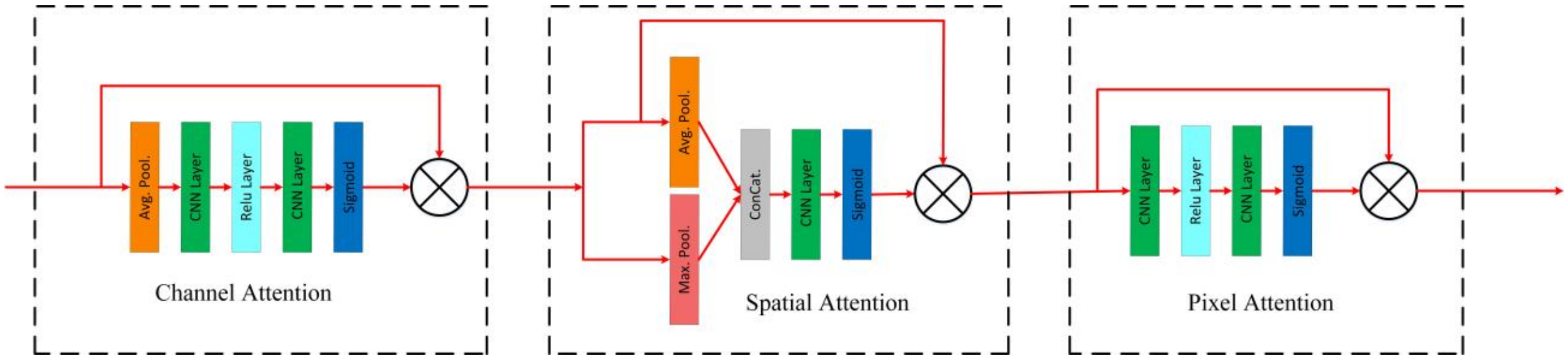

Fig. 4: Visualization of the channel, spatial, and pixel attention mechanisms in the CBSM model.

Based on the Fig. 4, the formulas of the mentioned attention mechanisms are shown as following.

$$M_C = \sigma\left(CNN_2\left(ReLU\left(CNN_1\left(GP_{avg}(x)\right)\right)\right)\right), \quad (1)$$

$$M_P = \sigma\left(CNN_2\left(ReLU\left(CNN_1(x)\right)\right)\right), \quad (2)$$

$$M_S = \sigma\left(CNN\left(concat\left(GP_{max}(x), GP_{avg}(x)\right)\right)\right), \quad (3)$$

where $x$ is the input to the attention mechanism, $GP_{avg}$ and $GP_{max}$ represent the global average pooling and global max pooling layer that reduces spatial dimensions and retains channel information. $CNN_1$ and $CNN_2$ are the convolutional neural network layers for learning channel-wise dependencies. $ReLU$ is the Rectified Linear Unit function. $\sigma$ is the Sigmoid function that normalizes the output to a range between 0 and 1.

By applying the CBSM after each down-sampling (in the encoder) and up-sampling (in the decoder) operation in the U-Net, the network can focus on the most important features at each level. This is because the CBSM refines the feature maps by giving more importance to the salient features in both the channel and spatial dimensions, and now with the addition of the pixel attention mechanism, at the pixel level as well. This could potentially improve the performance of the U-Net, especially when dealing with limited sample sizes, as the CBSM allows the network to make better use of the available data by focusing on the most informative parts of the input images. Figure 5 shows the U-Net architecture with CBSM applied for lung segmentation using chest X-ray imagery.

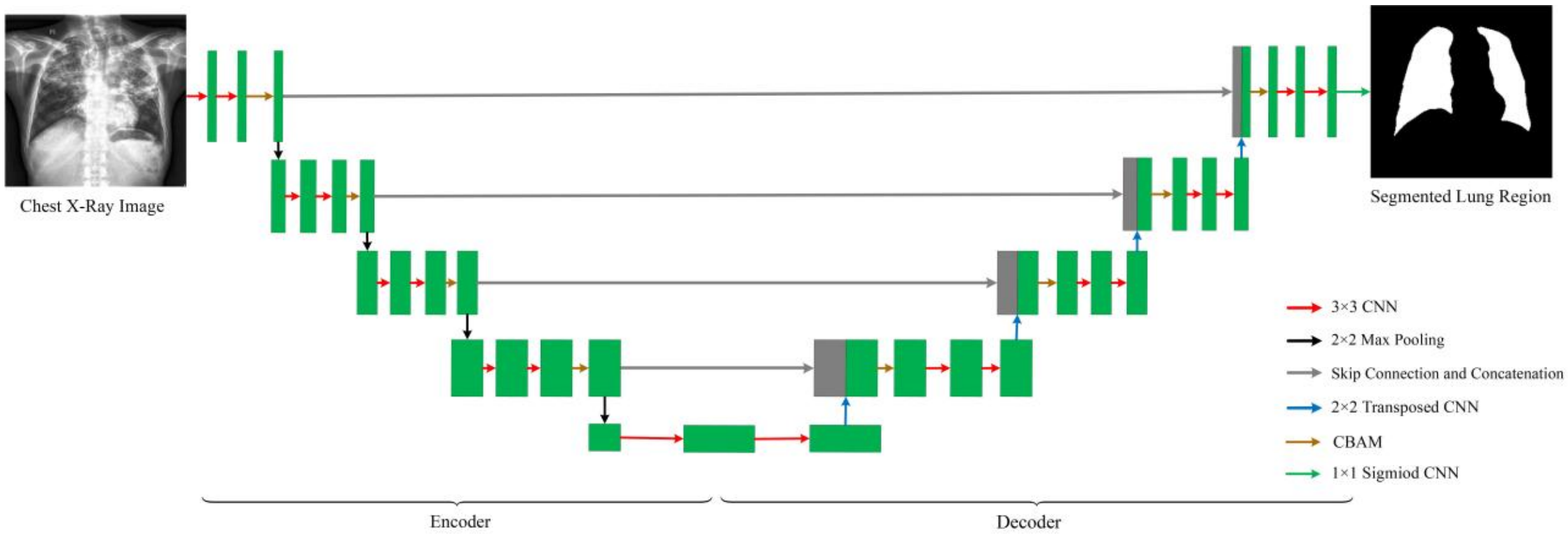

Fig. 5: U-Net architecture with CBSM applied for lung segmentation using chest X-ray imagery.

## 4) Simulations

In this section, we evaluate the effectiveness of the proposed method for chest X-ray segmentation using metrics such as the Dice similarity coefficient and Jaccard index. We also compare the ground truth with segmentation masks. Additionally, we assess segmentation performance using various metrics, including precision, recall, and accuracy.

### 4.1) Effectiveness of the Proposed Method using Dice similarity coefficient and Jaccard index

Semantic segmentation, also referred to as pixel-based classification, is a pivotal technique in which each pixel of an image is ascribed to a distinct category or class. This methodology is quintessential across a multitude of disciplines, encompassing medical imaging for the delineation of tissues, remote sensing for the classification of land cover, and autonomous driving for the interpretation of road scenarios. The primary objective of semantic segmentation is the allocation of a label to every individual pixel, ensuring that pixels bearing identical labels exhibit shared attributes. The evaluation of model performance within this context employs the Jaccard index and the Dice coefficient, which are critical metrics for assessing the accuracy of the segmentation. The Jaccard index and the Dice coefficient are integral metrics in the domain of semantic segmentation. They are predicated on the foundational elements of true positives (TP), false positives (FP), false negatives (FN), and true negatives (TN). True positives and true negatives refer to the tuberculosis images correctly identified as tuberculosis and the normal images correctly identified as normal, respectively. Conversely, false positives are normal images incorrectly identified as tuberculosis, and false negatives are tuberculosis images incorrectly identified as normal. The Jaccard index, also recognized as the Intersection over Union (IoU), quantifies the congruence between the predicted labels and the ground truth by calculating the ratio of the intersection to the union of the predicted and actual labels. Mathematically, it is represented as:

$$IoU = \frac{TP}{TP + FP + FN} \tag{4}$$

Conversely, the Dice coefficient, synonymous with the Dice similarity coefficient, gauges the overlap between two samples. It is computed as twice the intersection of the predicted and true labels, divided by the sum of the counts of both labels. The formula is expressed as:

$$Dice = \frac{2 \times TP}{2 \times TP + FP + FN} \tag{5}$$

These metrics are particularly advantageous in semantic segmentation for their ability to measure the extent of overlap between the segmentation predictions and the ground truth. The Dice Similarity Coefficient and the Jaccard Index using the are depicted in Figures 6 and 7.

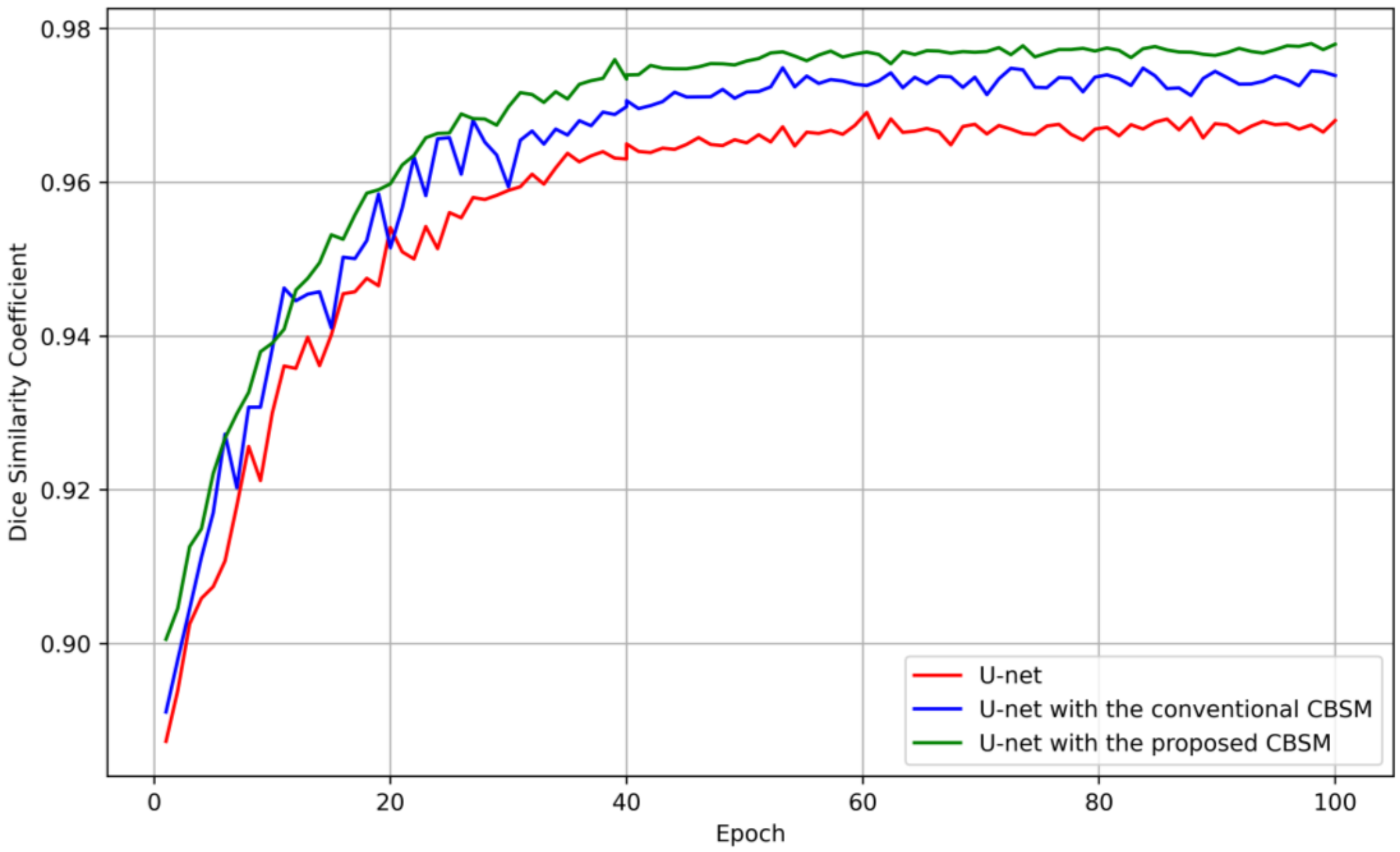

Fig. 6: Dice similarity coefficient using by U-net, U-net with the conventional CBSM[29], and U-net with the proposed CBSM.

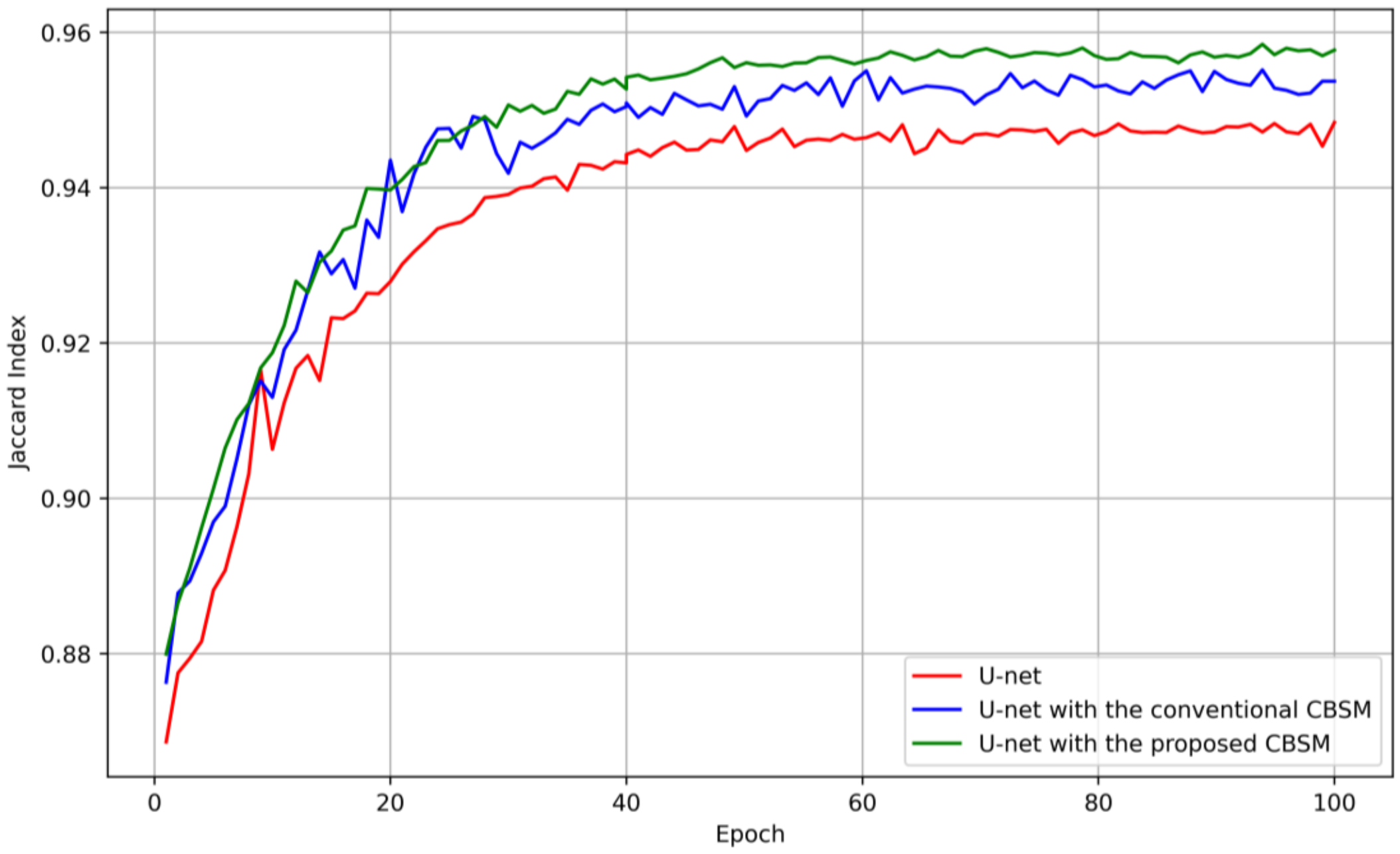

Fig. 7: Jaccard index or IoU using by U-net, U-net with the conventional CBSM [29], and U-net with the proposed CBSM.

The figures in question presumably depict the comparative performance of three distinct U-net architectures tailored for the segmentation of chest X-ray images. These performances are quantified using the Dice similarity coefficient and the Jaccard index. The superior performance of the U-net model augmented with the Proposed CBSM can be attributed to the following reasons:

1. **U-net without CBSM**: This is the baseline model that does not include any attention mechanisms. It likely performs the least effectively because it treats all channels and spatial locations equally, without focusing on the most informative features.

2. **U-net with Conventional CBSM**: This model incorporates channel and spatial attention mechanisms. The channel attention allows the model to weigh channels differently, prioritizing those that are more informative for the task. The spatial attention mechanism enables the model to focus on specific areas within the image that are more relevant to the segmentation task. These mechanisms help the model to ignore irrelevant information and concentrate on important features, leading to better performance than the baseline U-net.

3. **U-net with Proposed CBSM**: This model includes all three attention mechanisms: channel, spatial, and pixel attention. The addition of pixel attention allows the model to focus on the most informative pixels, which is particularly useful for fine-grained segmentation tasks. This level of detail enables the model to make more accurate predictions at the pixel level, resulting in a higher Dice similarity coefficient compared to the U-net with only channel and spatial attention.

The graph likely shows that the U-net with the proposed CBSM achieves the highest Dice similarity and Jaccard index coefficients across epochs, indicating that it outperforms the other two models. The inclusion of pixel attention, along with channel and spatial attention, provides a more nuanced understanding of the image, leading to superior segmentation results. This suggests that the proposed CBSM is more adept at capturing global contextual information and focusing the network's attention on specific, relevant regions of the chest X-ray images.

### 4.2) Comparison between the ground truth and Segmentation masks

Visual comparison between automated segmentation masks and manually annotated ground truth serves several critical purposes. It assesses accuracy, validates quantitative metrics, identifies algorithmic challenges, supports clinical decision-making, and enhances education and communication within the medical imaging field. The Fig. 8 shows a comparative display of chest X-ray segmentation results using different U-net architectures.

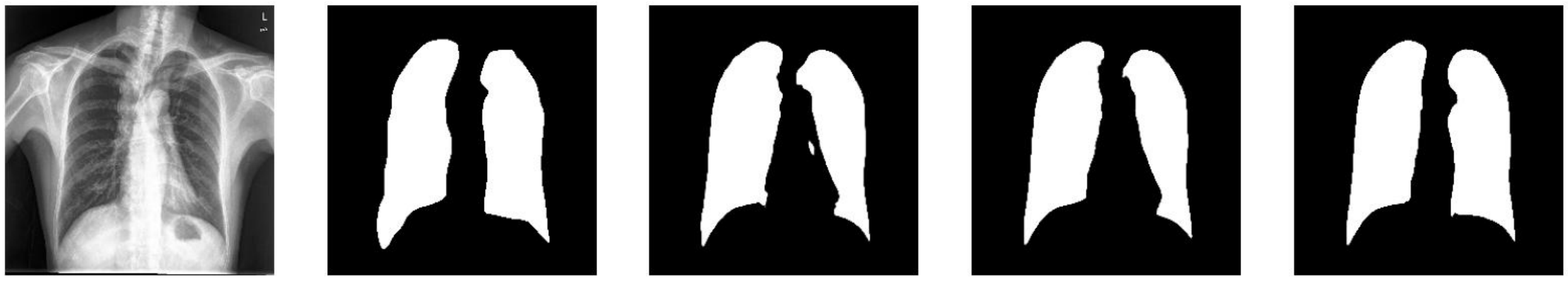

A B C D E

(1)

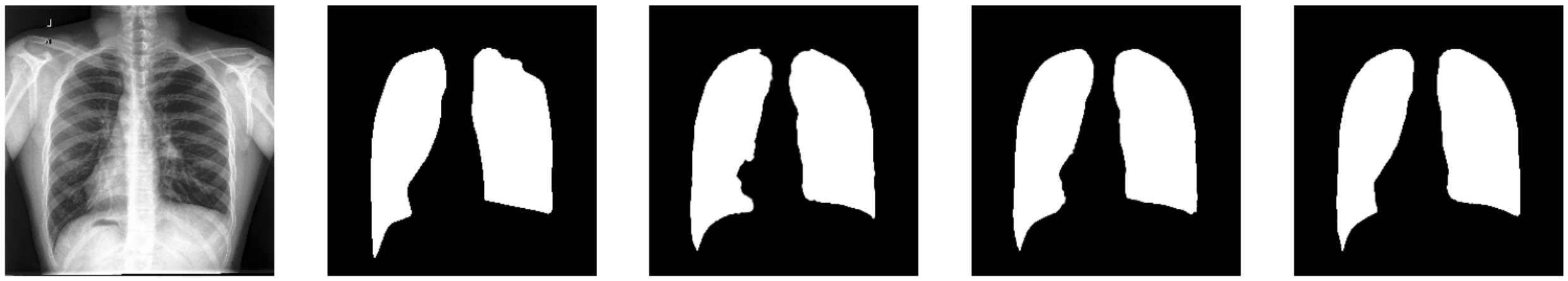

A B C D E

(2)

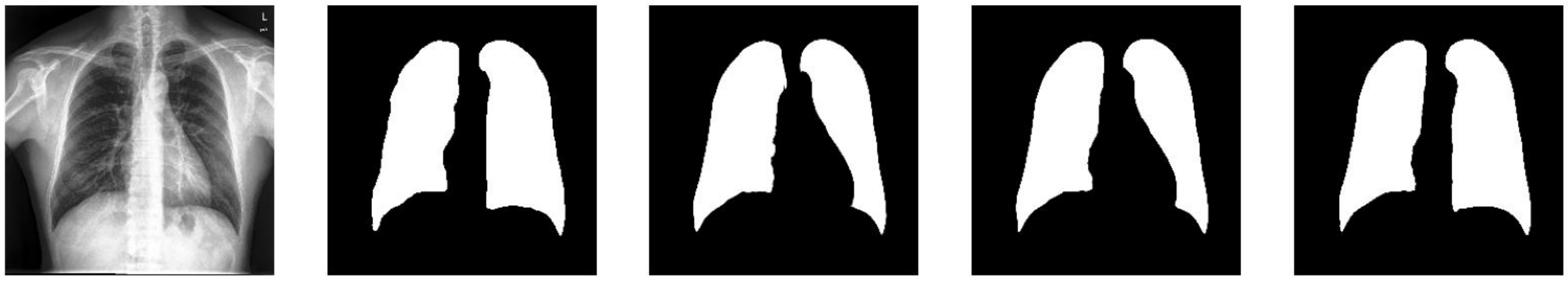

A B C D E

(3)

Fig.8: Display of three samples: Chest x-ray Image (A), the reference lung mask of the input chest X-ray (ground truth) (B), the segmented region generated by U-net (C), U-net with the conventional CBSM (D), and U-net with the proposed CBSM (E).

The progression from panels C to E demonstrates the incremental improvements in segmentation accuracy as more sophisticated attention mechanisms are incorporated into the U-net architecture based on the mentioned reasons. The proposed CBSM's inclusion of pixel attention, in addition to channel and spatial attention, allows for a more precise and detailed analysis of the chest X-ray images, leading to the most accurate segmentation results.

### 4.3) Segmentation Performance Evaluation Using Various Metrics

In this subsection, we embark on a comprehensive exploration of other performance indices for the effectiveness of our proposed segmentation method. These metrics are calculated using the following formulas:

- **Accuracy**: The proportion of true results (both true positives and true negatives) among the total number of cases examined.

$$Accuracy = \frac{TP + TN + FP + FN}{TP + FN} \tag{6}$$

- **Sensitivity (Recall)**: The proportion of actual positive cases which are correctly identified.

$$Recall \; = \frac{TP}{TP + FN} \tag{7}$$

- **Specificity**: The proportion of actual negative cases which are correctly identified.

$$Specificity = \frac{TN}{FP + TN} \tag{8}$$

- **Precision**: The proportion of positive identifications that were actually correct.

$$Precision \; = \frac{TP}{TP + FP}. \tag{9}$$

- **F1 Score**: The harmonic mean of precision and sensitivity, providing a balance between the two in cases where one may be more important than the other.

$$F1 - score \;\; = 2\, \frac{Precision \; \times Recall}{Precision \; + Recall}\,. \tag{10}$$

The metrics presented collectively offer a detailed assessment of the deep learning model's efficacy in pixel-level classification tasks within chest X-ray (CXR) imagery. Table 1 delineates the aforementioned performance indices.

**Table 1.** The performance indices using U-net, U-net with the conventional CBSM [29], and U-net with the proposed CBSM

| Method | Accuracy (%) | Recall (%) | Specificity (%) | Precision (%) | F1-score (%) |
|---|---|---|---|---|---|
| **U-net** | 96.2 | 95.30 | 93.54 | 96.68 | 95.98 |
| **U-net with the conventional CBSM** [29] | 97.8 | 95.57 | 95.81 | 97.14 | 96.34 |
| **U-net with the proposed CBSM** | 98.8 | 97.50 | 97.64 | 97.68 | 97.58 |

The results in Table 1 can be interpreted as follows:

- **U-net**: The standard U-net model shows high effectiveness with an accuracy of **96.2%**, a recall of **95.30%**, and a precision of **96.68%**, leading to an F1-score of **95.98%**. Its specificity is **93.54%**, indicating it is slightly less effective at correctly identifying non-relevant instances compared to relevant ones.
- **U-net with the conventional CBSM**: The addition of channel and spatial attention mechanisms (CBSM) improves the model's performance across all metrics, with notable increases in specificity and precision, resulting in an F1-score of **96.34%**.
- **U-net with the proposed CBSM**: Incorporating the proposed CBSM, which likely includes additional pixel attention mechanisms, further enhances the model's performance, achieving the highest accuracy of **98.8%**, recall of **97.50%**, and an F1-score of **97.58%**. This model also shows a high specificity of **97.64%**, indicating a strong ability to correctly identify non-relevant instances.

The incremental improvements in these metrics suggest that the proposed CBSM provides a more nuanced and detailed analysis of CXR images, leading to more accurate segmentation results. The reference to [29] suggests that the conventional CBSM model was likely detailed in their work, and the proposed CBSM is an advancement over this model.

## 5) Conclusion

The research presents an innovative approach to lung segmentation in chest X-ray imagery, enhancing the U-net architecture through the incorporation of a CBAM. This module amalgamates three distinct attention mechanisms—channel attention, spatial attention, and pixel attention—refining the model's structural framework for improved performance. The channel, spatial, and pixel attention mechanisms collectively enhance a model's segmentation accuracy. The channel attention prioritizes significant channels, the spatial attention targets important spatial areas, and the pixel attention zeroes in on the most informative pixels, all contributing to a more precise and detailed segmentation output. The resulting improvements in feature representation and segmentation accuracy are not merely theoretical but have been empirically validated through rigorous performance assessments. The utilization of metrics such as the Dice coefficient and Jaccard similarity index has provided a quantifiable measure of the method's superiority over traditional models. Moreover, the comparative analysis of pixel classification metrics across different U-net architectures for chest X-ray image segmentation reveals a clear trend of performance enhancement with the integration of attention mechanisms. In conclusion, the integration of the proposed CBAM with the U-net architecture represents a significant leap forward in the domain of medical image analysis.

Future investigations aimed at refining the attention mechanisms for targeted lung pathology features have the potential to significantly enhance the precision of segmentation outcomes.

Additionally, the exploration of integrating a U-net architecture augmented with a CBAM into diverse imaging modalities, including CT and MRI scans, holds promise for the development of a more holistic diagnostic instrument.

## 6) Conflicts of Interest Declaration

All authors declare that they have no financial or non-financial conflicts of interest related to the work presented in this manuscript.

## 7) Reference

[1] D. I. Morís, J. de Moura, S. Aslani, J. Jacob, J. Novo, and M. Ortega, "Multi-task localization of the hemidiaphragms and lung segmentation in portable chest X-ray images of COVID-19 patients," *Digit. Heal.*, vol. 10, p. 20552076231225852, 2024, doi: 10.1177/20552076231225853.

[2] D. N. Kumar and M. K. Joseph, "Fused Feature Vector and Dual FCM for Lung Segmentation from Chest X-Ray Images.," *Int. J. Intell. Eng. Syst.*, vol. 17, no. 2, 2024, doi: 10.22266/ijies2024.0430.38.

[3] M. A. L. Khaniki, M. Mirzaeibonehkhater, and M. Manthouri, "Enhancing Pneumonia Detection using Vision Transformer with Dynamic Mapping Re-Attention Mechanism," in *2023 13th International Conference on Computer and Knowledge Engineering (ICCKE)*, IEEE, 2023, pp. 144–149. doi: 10.1109/ICCKE60553.2023.10326313.

[4] H. Ajami, M. K. Nigjeh, and S. E. Umbaugh, "Unsupervised white matter lesion identification in multiple sclerosis ( MS ) using MRI segmentation and pattern classification : a novel approach with CVIPtools," vol. 12674, pp. 1–6, 2023, doi: 10.1117/12.2688268.

[5] A. Birashk, J. K. Kordestani, and M. R. Meybodi, "Cellular teaching-learning-based optimization approach for dynamic multi-objective problems," *Knowledge-Based Syst.*, vol. 141, pp. 148–177, 2018, doi: 10.1016/j.knosys.2017.11.016.

[6] A. Samii, H. Karami, H. Ghazvinian, A. Safari, and Y. D. Ajirlou, "Comparison of DEEP-

LSTM and MLP Models in Estimation of Evaporation Pan for Arid Regions.," *J. Soft Comput. Civ. Eng.*, vol. 7, no. 2, 2023.

[7] V. Monjezi, A. Trivedi, G. Tan, and S. Tizpaz-Niari, "Information-theoretic testing and debugging of fairness defects in deep neural networks," in *2023 IEEE/ACM 45th International Conference on Software Engineering (ICSE)*, IEEE, 2023, pp. 1571–1582. doi: 10.1109/ICSE48619.2023.00136.

[8] M. Salehi, N. Javadpour, B. Beisner, M. Sanaei, and S. B. Gilbert, "Innovative Cybersickness Detection: Exploring Head Movement Patterns in Virtual Reality," *arXiv Prepr. arXiv2402.02725*, 2024, doi: 10.48550/arXiv.2402.02725.

[9] M. Sanaei, S. B. Gilbert, N. Javadpour, H. Sabouni, M. C. Dorneich, and J. W. Kelly, "The Correlations of Scene Complexity, Workload, Presence, and Cybersickness in a Task-Based VR Game," *arXiv Prepr. arXiv2403.19019*, 2024, doi: 10.48550/arXiv.2403.19019.

[10] N. Devroye *et al.*, "Interpreting deep-learned error-correcting codes," in *2022 IEEE International Symposium on Information Theory (ISIT)*, IEEE, 2022, pp. 2457–2462.

[11] M. K. Nigjeh, H. Ajami, and S. E. Umbaugh, "Automated classification of white matter lesions in multiple sclerosis patients ' MRI images using gray level enhancement and deep learning," vol. 12674, pp. 1–6, 2023, doi: 10.1117/12.2688269.

[12] M. Mehrnia, M. Elbayumi, and M. S. M. Elbaz, "Assessing Foundational Medical'Segment Anything'(Med-SAM1, Med-SAM2) Deep Learning Models for Left Atrial Segmentation in 3D LGE MRI," *arXiv Prepr. arXiv2411.05963*, 2024.

[13] S. N. Nia and F. Y. Shih, "Medical X-Ray Image Enhancement Using Global Contrast-Limited Adaptive Histogram Equalization," *arXiv Prepr. arXiv2411.01373*, 2024.

[14] F. Azari *et al.*, "Elucidating the high compliance mechanism by which the urinary bladder fills under low pressures," *arXiv Prepr. arXiv2501.10312*, 2025.

[15] O. Ronneberger, P. Fischer, and T. Brox, "U-net: Convolutional networks for biomedical image segmentation," in *Medical image computing and computer-assisted intervention–MICCAI 2015: 18th international conference, Munich, Germany, October 5-9, 2015,*

*proceedings, part III 18*, Springer, 2015, pp. 234–241. doi: 10.1007/978-3-319-24574-4_28.

[16] Z. Zhou, M. M. Rahman Siddiquee, N. Tajbakhsh, and J. Liang, "Unet++: A nested u-net architecture for medical image segmentation," in *Deep Learning in Medical Image Analysis and Multimodal Learning for Clinical Decision Support: 4th International Workshop, DLMIA 2018, and 8th International Workshop, ML-CDS 2018, Held in Conjunction with MICCAI 2018, Granada, Spain, September 20, 2018, P*, Springer, 2018, pp. 3–11. doi: 10.1007/978-3-030-00889-5_1.

[17] T. Agrawal and P. Choudhary, "ReSE-Net: Enhanced UNet architecture for lung segmentation in chest radiography images," *Comput. Intell.*, vol. 39, no. 3, pp. 456–477, 2023, doi: 10.1111/coin.12575.

[18] V.-L. Le and O. Saut, "RRc-UNet 3D for lung tumor segmentation from CT scans of Non-Small Cell Lung Cancer patients," in *Proceedings of the IEEE/CVF International Conference on Computer Vision*, 2023, pp. 2316–2325.

[19] M. A. L. Khaniki and M. Manthouri, "Enhancing Price Prediction in Cryptocurrency Using Transformer Neural Network and Technical Indicators," *arXiv Prepr. arXiv2403.03606*, 2024, doi: 10.48550/arXiv.2403.03606.

[20] M. A. Labbaf-Khaniki, M. Manthouri, and H. Ajami, "Twin Transformer using Gated Dynamic Learnable Attention mechanism for Fault Detection and Diagnosis in the Tennessee Eastman Process," *arXiv Prepr. arXiv2403.10842*, 2024, doi: 10.48550/arXiv.2403.10842.

[21] H. Irani and V. Metsis, "Positional Encoding in Transformer-Based Time Series Models: A Survey," *arXiv Prepr. arXiv2502.12370*, 2025.

[22] M. Akyash, A. Zafari, and N. M. Nasrabadi, "Trading-off Mutual Information on Feature Aggregation for Face Recognition," *arXiv Prepr. arXiv2309.13137*, 2023, doi: 10.48550/arXiv.2309.13137.

[23] M. A. Khalil, M. Sadeghiamirshahidi, R. M. Joeckel, F. M. Santos, and A. Riahi, "Mapping a hazardous abandoned gypsum mine using self-potential, electrical resistivity

tomography, and Frequency Domain Electromagnetic methods," *J. Appl. Geophys.*, vol. 205, p. 104771, 2022, doi: 10.1016/j.jappgeo.2022.104771.

[24] Y. Zhang *et al.*, "CSANet: Channel and spatial mixed attention CNN for pedestrian detection," *IEEE Access*, vol. 8, pp. 76243–76252, 2020, doi: 10.1109/ACCESS.2020.2986476.

[25] H. Guo, J. Liu, J. Yang, Z. Xiao, and Z. Wu, "Deep collaborative attention network for hyperspectral image classification by combining 2-D CNN and 3-D CNN," *IEEE J. Sel. Top. Appl. Earth Obs. Remote Sens.*, vol. 13, pp. 4789–4802, 2020, doi: 10.1109/JSTARS.2020.3016739.

[26] H. Jia *et al.*, "A model combining multi branch spectral-temporal CNN, Efficient Channel attention, and LightGBM for MI-BCI classification," *IEEE Trans. Neural Syst. Rehabil. Eng.*, vol. 31, pp. 1311–1320, 2023, doi: 10.1109/TNSRE.2023.3243992.

[27] I. Aboussaleh, J. Riffi, K. El Fazazy, M. A. Mahraz, and H. Tairi, "Efficient U-Net architecture with multiple encoders and attention mechanism decoders for brain tumor segmentation," *Diagnostics*, vol. 13, no. 5, p. 872, 2023, doi: 10.3390/diagnostics13050872.

[28] A. Amer, T. Lambrou, and X. Ye, "MDA-unet: a multi-scale dilated attention U-net for medical image segmentation," *Appl. Sci.*, vol. 12, no. 7, p. 3676, 2022, doi: 10.3390/app12073676.

[29] G. Gaál, B. Maga, and A. Lukács, "Attention U-net based adversarial architectures for chest X-ray lung segmentation," *CEUR Workshop Proc.*, vol. 2692, pp. 1–7, 2020, doi: 10.48550/arXiv.2003.10304.

[30] T. Rahman *et al.*, "Reliable tuberculosis detection using chest X-ray with deep learning, segmentation and visualization," *IEEE Access*, vol. 8, pp. 191586–191601, 2020, doi: 10.1109/ACCESS.2020.3031384.